\documentclass[10pt, fleqn]{article}
\usepackage[utf8]{inputenc}
\DeclareUnicodeCharacter{2212}{$-$}
\usepackage{graphicx}
\usepackage[fleqn]{amsmath}
\usepackage[version=4]{mhchem}
\usepackage{siunitx}
\usepackage{longtable,tabularx}
\usepackage{nccmath}
\usepackage{amsmath}
\usepackage[superscript]{cite}
\usepackage[margin=1in]{geometry}
\usepackage[toc,page]{appendix}
\usepackage{setspace}
\usepackage{amssymb}
\usepackage{tabu}
\usepackage{listings}
\usepackage{afterpage}
\usepackage{titlesec}
\usepackage[labelfont=bf]{caption}
\usepackage{fancyhdr}
\usepackage{pgf}
\usepackage{float}
\usepackage{subcaption}
\usepackage{aas_macros}
\usepackage{cleveref}
\usepackage{textcomp}
\usepackage{pict2e}
\usepackage{wasysym}
\usepackage{multicol}
\usepackage{lipsum}
\usepackage{indentfirst}
\newenvironment{Figure}
  {\par\medskip\noindent\minipage{\linewidth}}
  {\endminipage\par\medskip}

\titleformat{\section}
{\normalfont\normalsize\bfseries}{\thesection}{0.5em}{}
\titleformat{\subsection}
{\normalfont\normalsize\bfseries\em}{\thesubsection}{0.5em}{}
\titleformat{\subsubsection}
{\normalfont\normalsize\bfseries\em}{\thesubsubsection}{0.5em}{}
\begin{document}

\begin{flushright}
\Large 

\textbf{SSC21-VI-02}
\end{flushright}
\begin{centering}      
\large 

\textbf{The Pandora SmallSat: Multiwavelength Characterization of Exoplanets and their Host Stars}\\
\vspace{0.5cm}
\normalsize 

{Elisa~V.~Quintana}, Knicole~D.~Colón, Gregory Mosby, Joshua E. Schlieder\\
{NASA Goddard Space Flight Center}\\
{8800 Greenbelt Road, Greenbelt, MD 20771}; 301-286-0851\\ 
{elisa.quintana@nasa.gov}\\ 
\vspace{0.5cm}
{Pete~Supsinskas}, Jordan~Karburn\\
{Lawrence Livermore National Laboratory}\\
{Livermore, California 94550}\\
\vspace{0.5cm}
{Jessie~L.~Dotson}, Thomas P. Greene, Christina Hedges\\
{NASA Ames Research Center}\\
{Moffett Field, CA 94035}\\
\vspace{0.5cm}
Dániel~Apai\\
{Steward Observatory, The University of Arizona}\\
{933 N. Cherry Avenue, Tucson, AZ 85721}\\
\vspace{0.5cm}
Thomas~Barclay\\
{University of Maryland, Baltimore County and NASA Goddard}\\
{1000 Hilltop Cir, Baltimore, MD 21250}\\
\vspace{0.5cm}
Jessie~L.~Christiansen\\
{Caltech/IPAC, 1200 E. California Blvd. Pasadena, CA 91125}\\
\vspace{0.5cm}
{Néstor Espinoza}, Susan E. Mullally\\
{Space Telescope Science Institute}\\
{3700 San Martin Drive, Baltimore, MD, 21218}\\
\vspace{0.5cm}
Emily A. Gilbert\\
{Department of Astronomy and Astrophysics, University of Chicago}\\
{5640 S. Ellis Ave, Chicago, IL 60637}\\
\vspace{0.5cm}
{Kelsey Hoffman}, Veselin B. Kostov\\
{SETI Institute}\\
{189 Bernardo Ave, Suite 200, Mountain View, CA 94043}\\
\vspace{0.5cm}
{Nikole K. Lewis}, Trevor O. Foote\\
{Carl Sagan Institute, Cornell University}\\
{Space Science Institute 312, 14850 Ithaca, NY}\\
\vspace{0.5cm}
{James Mason}, Allison Youngblood\\
{Laboratory for Atmospheric and Space Physics, University of Colorado}\\
{1234 Innovation Dr, Boulder, CO 80303}\\
\vspace{0.5cm}
Brett M.~Morris\\
{Center for Space and Habitability, University of Bern}\\
{Gesellschaftsstrasse 6, 3012 Bern, Switzerland}\\
\vspace{0.5cm}
Elisabeth R. Newton\\
{Department of Physics and Astronomy, Dartmouth College}\\
{Hanover, NH 03755}\\
\vspace{0.5cm}
Joshua Pepper\\
{Physics Department, Lehigh University}\\
{Bethlehem, PA 18015}\\
\vspace{0.5cm}
Benjamin V.\ Rackham\\
{Department of Earth, Atmospheric and Planetary Sciences, Massachusetts Institute of Technology}\\
{77 Massachusetts Avenue, Cambridge, MA 02139}\\
\vspace{0.5cm}
Jason~F.~Rowe\\
{Department of Physics and Astronomy, Bishops University}\\
{2600 College St, Sherbrooke, QC J1M 1Z7, Canada}\\
\vspace{0.5cm}
Kevin Stevenson\\
{Johns Hopkins Applied Physics Laboratory}\\
{11100 Johns Hopkins Road, Laurel, Maryland 20723}\\
\vspace{0.5cm}

\vspace{0.5cm}
\centerline{\textbf{ABSTRACT}}
\vspace{0.3cm}
\end{centering}
Pandora is a SmallSat mission designed to study the atmospheres of exoplanets, and was selected as part of NASA's Astrophysics Pioneers Program. Transmission spectroscopy of transiting exoplanets provides our best opportunity to identify the makeup of planetary atmospheres in the coming decade. Stellar brightness variations due to star spots, however, can impact these measurements and contaminate the observed spectra. Pandora's goal is to disentangle star and planet signals in transmission spectra to reliably determine exoplanet atmosphere compositions. Pandora will collect long-duration photometric observations with a visible-light channel and simultaneous spectra with a near-IR channel. The broad-wavelength coverage will provide constraints on the spot and faculae covering fractions of low-mass exoplanet host stars and the impact of these active regions on exoplanetary transmission spectra. Pandora will subsequently identify exoplanets with hydrogen- or water-dominated atmospheres, and robustly determine which planets are covered by clouds and hazes. Pandora will observe at least 20 exoplanets with sizes ranging from Earth-size to Jupiter-size and host stars spanning mid-K to late-M spectral types. The project is made possible by leveraging investments in other projects, including an all-aluminum 0.45-meter Cassegrain telescope design, and a NIR sensor chip assembly from the James Webb Space Telescope. The mission will last five years from initial formulation to closeout, with one-year of science operations. Launch is planned for the mid-2020s as a secondary payload in Sun-synchronous low-Earth orbit. By design, Pandora has a diverse team, with over half of the mission leadership roles filled by early career scientists and engineers, demonstrating the high value of SmallSats for developing the next generation of space mission leaders.    

%

\begin{multicols*}{2}

\section{Introduction}

\begin{figure*}
\centering
\includegraphics[width=\textwidth]{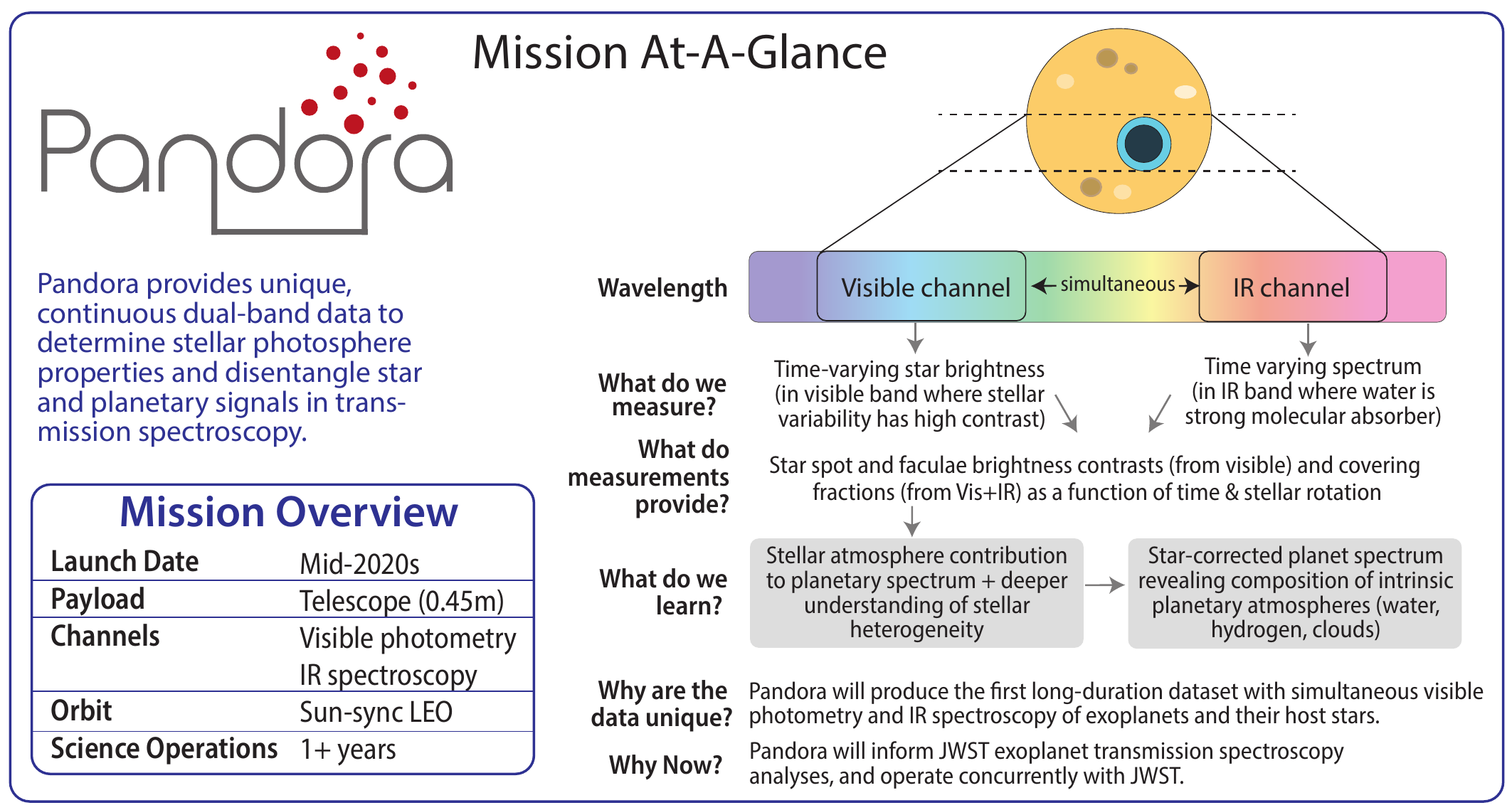}
\captionof{figure}{Pandora Mission-At-A-Glance.}
\label{fig:minifactsheet}
\end{figure*}

Exoplanet transmission spectroscopy is a proven technique used to probe the atmospheres of distant exoplanets and is a primary science case for the upcoming James Webb Space Telescope (JWST). As we move into an era of probing the atmospheres of increasingly smaller planets---the most accessible of which orbit small, active stars---with increasingly higher precision, the importance of understanding the effects of stellar variability on transmission spectroscopy measurements grows.  Pandora is a SmallSat mission designed to disentangle star and planet signals in transmission spectra of transiting exoplanets in order to obtain robust measurements of planetary atmospheres. Figure~\ref{fig:minifactsheet} provides a high-level overview of the mission. Pandora will obtain long-duration, simultaneous multiwavelength---visible and near-infrared (NIR)---observations of 20 exoplanets and their host stars to quantify and correct for stellar spectral contamination in transmission spectra. Pandora data will subsequently be used to identify which of these planets have hydrogen- or water-dominated atmospheres, and which planets appear to be covered by clouds and hazes.

Pandora is one of four concepts selected as part of the first NASA Astrophysics Pioneers Program. Pioneers missions are designed to enable compelling astrophysics science at a lower cost than missions in NASA's Explorers Program, and have a 5-year performance period from selection to closeout. Pandora's launch is planned for the mid-2020s as a secondary payload, and the baseline plan is for one year of science operations although an extended mission is possible that could encompass additional ancillary science opportunities. The Pandora mission is co-led by NASA Goddard Space Flight Center (GSFC) and Lawrence Livermore National Laboratory (LLNL), in collaboration with NASA Ames Research Center (ARC) and more than a dozen universities and research institutes. 

\section{Science Rationale}
Thousands of planets orbiting stars outside our solar system have been discovered, revealing a dazzling diversity of planetary systems. This leads to a natural variety of planetary atmospheres. The unique spectral fingerprints of atmospheres encode information about planet formation, composition, and evolution and provide access to unexplored super-Earths and sub-Neptunes not seen in our solar system. Atmospheres are also a key observable in identifying which exoplanets may have habitable climatic states. Exoplanet transmission spectroscopy is a proven technique to probe the atmospheres of exoplanets that cross in front of---or transit---their star \cite{Seager2000}. During a transit, starlight filters through the planet’s atmosphere and the differential measurement of the observed in-transit spectra (planet$+$star) and the out-of-transit spectra (star-only) should yield the spectra of the planet's atmosphere. 

During a planet's transit, we observe a decrease in flux from a star due to a planet blocking a small fraction of a star's light. The key to exoplanet transmission spectroscopy is that the transit depth is \textit{wavelength dependent}. For a given wavelength, some atoms and molecules in the planet's atmosphere absorb incident light, while others allow light to pass through. Greater absorption causes the disk of the planet to appear more opaque and larger, leading to deeper observed transits (as transit depths are proportional to the planet-to-star radius ratio). The wavelength-resolved observation of transit depths yields a transmission spectrum, a unique spectral fingerprint which can then be compared to models to infer the planet's atmospheric constituents. The Hubble Space Telescope (HST) and ground-based observatories have already been used to perform transmission spectroscopy to identify the presence of molecules, gases, and clouds in the atmospheres of giant exoplanets \cite{Sing2016}. Exoplanetary transmission spectroscopy with JWST will probe the atmospheres of planets as small as Earth with unprecedented precision \cite{Greene2016,Greene2019}. 

Underpinning transmission spectroscopy measurements is an approximation that the planet’s host star---the light source illuminating the planet’s atmosphere---can be treated as a uniform light source. In reality, many stars are magnetically active, producing star spots (regions that radiate less light than the undisturbed photosphere) and faculae (regions that radiate more light) that cause the emergent light to vary across the stellar disk \cite{Baliunas1995}. These changes in surface brightness also vary with time, as spots and faculae can evolve spatially, and the observed spot/faculae configuration varies as the spots/faculae come into and out of view as a star rotates. The observed stellar disk spot/faculae configuration can therefore be different for each transit of the same planet, and for individual planets in a multi-planet system. This makes interpreting a single exoplanet transit challenging in the presence of stellar inhomogeneity and introduces an astrophysical systematic noise floor when combining multiple transits. 

In addition, the light source filtering through the planet’s atmosphere comes from the transit chord, the projected area of the star that the planet occults (Figure~\ref{fig:tlse}), which is only a fraction of the stellar disk (the assumed light source). Spatial inhomogeneities in the occulted stellar photosphere (spot crossings) are more easily identified since they can be imprinted in the transit light curve, albeit they are still challenging to quantify \cite{Dittmann2009,Sing2011,Deming2011,Sanchis2011,Morris2017}. Inhomogeneities in the unocculted stellar photosphere, however, are more difficult to both identify and quantify. The spectral difference between the assumed (disk-integrated) light source and the actual (chord-integrated) light source will be imprinted in the observed transmission spectrum, leading to stellar spectral contamination in the deduced planetary spectrum, a phenomenon known as the transit light source effect \cite{Rackham2018}. Stellar contamination can introduce positive or negative features in an observed spectrum that can mimic or suppress features in the intrinsic planetary spectrum \cite{McCullough2014,Rackham2017,Rackham2018,Apai2018,Espinoza2019,Rackham2019b,Rackham2019a,Wakeford2019}. This can lead to false detections of planetary atmospheric features, or can suppress features that could lead to an exciting discovery. 


The most accessible planets for atmospheric studies orbit low-mass stars (K and M dwarfs, stars smaller than the Sun), because the strength of the planetary atmospheric features scales inversely with the square of the star size. Planets transiting smaller stars thus produce more significant absorption features and can be better characterized. However, stellar spectral contamination is most problematic in this regime because small stars retain high levels of magnetic activity over billion-year timescales. Furthermore, rotation periods of low-mass stars can be short (hours-to-days), so spot variability timescales can be comparable to planet orbital periods and transit durations. These small, cool stars can also have strong molecular absorption features in their stellar atmospheres (e.g., those caused by water vapor), which can be mistaken for planetary atmosphere features if left uncorrected.  

\begin{Figure}
\includegraphics[width=\textwidth]{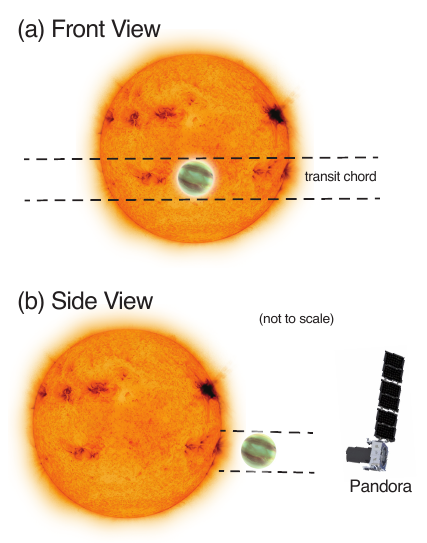}
\captionof{figure}{Most studies assume the stellar disk is the light source in exoplanet transmission spectroscopy. The actual light source filtering through a planet's atmosphere is the transit chord. Spectral features from the unocculted stellar regions contribute to stellar contamination of the transmission spectra, known as the transit light source effect \cite{Rackham2018}, which Pandora is designed to measure.}
\label{fig:tlse}
\end{Figure}

Stellar contamination in transmission spectra has been systematically underestimated in the exoplanet literature \cite{Apai2018}, resulting in several past HST results now being called into question \cite{McCullough2014,Pinhas2018,Zhang2018}. While stellar contamination of transit spectra has been identified as a concern for more than a decade \cite{Pont2008,Pont2013,Sing2011,Berta2012,McCullough2014,Deming2017}, with the advance of high-precision transmission spectroscopy its importance has skyrocketed. Today, stellar heterogeneity, and the resulting contamination in transmission spectra, is widely considered to be one of the limiting factors for HST and JWST transit spectroscopy and for the characterization of small planets. There is a broad consensus in the exoplanet community about the importance of this problem: for example, it was highlighted by the NAS Committee on Exoplanet Science Strategy \cite{NAP25187}, the HST-TESS Advisory Committee report (2019), the NASA Exoplanet Exploration Program (which launched a new Study Analysis Group to address the issue), and multiple, independent white papers submitted to the 2020 Astrophysics Decadal Survey \cite{Kowalski2019, Rackham2019b}. Yet, in spite of previous studies that sought solutions to stellar contamination, none could derive a universal solution, largely due to the limited data available. Pandora will provide the first dataset with simultaneous, multiband (visible and NIR), long-baseline observations of exoplanets and their host stars to address this key issue.

%

Considering that over 30 planets currently on the JWST Guaranteed Time Observations (GTO), Early Release Science (ERS), and General Observer (GO) target lists orbit stars smaller than the Sun, there is a critical need to observationally assess the impact of stellar contamination on exoplanet atmosphere spectra. Pandora's long-duration and broad-wavelength observations are unique and optimized to quantify stellar contamination from low-mass stars.

\section{Pandora Science Objectives}
Pandora's goal is to disentangle star and planet signals in transmission spectra to reliably determine exoplanet atmosphere compositions. Pandora has two science objectives in support of this goal which drive the science and mission requirements.

\subsection{Objective I:  Measure Host-star Spot Coverage}
 
\noindent Pandora's first objective is to determine the spot and faculae covering fractions of low-mass exoplanet host stars and the impact of these active regions on exoplanetary transmission spectra. Pandora will monitor stars using a visible monochromatic camera (wider than 450--650\,nm), which provides precise time series measurements of a star's brightness. A NIR channel (wider than 1000--1600 nm) will operate simultaneously with the visible channel to capture changes in spectral features that correlate with stellar photometric variability. The strength of Pandora's combined time-series photometry$+$spectra is that it will provide constraints on a star's absolute spot coverage.

Pandora's visible photometry and simultaneous NIR spectroscopy will provide the necessary spectral information to allow us to decompose the stellar spectrum into its constituent parts (spot, faculae, and quiescent photosphere) \cite{Zhang2018, Wakeford2019}. This can then be used to measure the spot/faculae contrasts and covering fractions during the planetary transit and provide upper limits on the stellar contamination signals that are present in the transmission spectra \cite{Rackham2018, Wakeford2019}. Because some of the star's surface is being blocked during the transit, Pandora's observations will directly constrain the sizes and contrasts of active regions within the transit chord \cite{Espinoza2019}. Pandora will combine both multiwavelength stellar monitoring and planet transit observations at the precisions required to break the degeneracy between active-region contrasts and covering fractions, determine full-disk and transit-chord active-region covering fractions, and robustly calculate stellar contamination signals.

\noindent Pandora's first objective will address:

Ia. \textbf{What are typical spot coverages of low-mass exoplanet host stars, and how do they vary with time?} Pandora will measure spot and faculae covering fractions and quantify stellar contamination as a function of time and stellar rotation phase.

Ib. \textbf{How do stellar properties (size, mass, temperature) correlate with contamination, and how does the impact of contamination change with planetary properties (size, mass, bulk density, orbital distance)?} Pandora’s notional target list includes planets as small as Earth and as large as Jupiter, with host stars spanning mid-K to late-M. Pandora will identify trends with stellar contamination and planet/star properties that will be used as inputs to models that generate and interpret exoplanetary transmission spectra.

Pandora will ultimately enable population-level insights into the stellar activity of K and M stars.


\subsection{Objective II: Identify Water- or Hydrogen-Dominated Atmospheres}

\noindent Pandora's second objective is to identify exoplanets with hydrogen- or water-dominated atmospheres, and determine which planets are likely covered by clouds and hazes. Pandora will obtain transmission spectra of 20 exoplanets at NIR wavelengths similar to those covered by HST's Wide Field Camera 3 G102 and G141 filters, where water is a known strong molecular absorber in both hydrogen- and water-dominated atmospheres. Jupiter-size planets typically have light, hydrogen-dominated atmospheres, while smaller planets are expected to have heavier atmospheres dominated by water or even heavier molecules.  


Pandora’s requirements are designed to deliver sufficient signal-to-noise ratio (SNR) to detect the atmospheric features due to water. 
A non-detection of spectral features consistent with H$_2$- or H$_2$O-dominated atmospheres implies that either clouds or hazes are obscuring atmospheric features, the atmosphere is dominated by heavier molecules than water, or the planet simply has no atmosphere at all. In any case, Pandora's multiwavelength observations will enable the spectral contamination to be removed from the planetary spectrum to provide a robust limit on the detection of atmospheric features. It will thus be possible to place rigorous upper limits on atmospheric composition even in the absence of a detection of spectral features, which can then be used as a driver for additional follow-up observations with other facilities like JWST. 

\noindent Pandora’s Objective II addresses: 

IIa. \textbf{How does the atmospheric composition of planets vary with radius/mass/bulk density, orbital distance, and host star properties?} Pandora’s notional target list includes a wide range of planet sizes and properties and stellar spectral types (Section~\ref{sec:targets}). 
Pandora will be sensitive to both lighter H$_2$-dominated and heavier H$_2$O-dominated atmospheres. 

IIb. \textbf{Which prior transmission spectroscopy observations yield the same atmospheric results after correcting for stellar contamination?} Pandora’s notional target list will include planets previously observed by HST, including the sub-Neptunes GJ\,436 b, GJ\,3470 b, and planets in the TRAPPIST-1 system. 

\subsection{Target Selection} \label{sec:targets}
To complete the science objectives, Pandora will target stars ranging from mid-K to late-M spectral type, and planets ranging from Earth-size to Jupiter-size with orbital periods up to about 25 days (corresponding to equilibrium temperatures of 200--1500 K). The targets include both rapidly and slowly rotating stars, and some stars have multiple planets. The wide parameter space of stars/planets enables Pandora to identify trends that yield predictions to aid target selection for future surveys. The Transiting Exoplanet Survey Satellite (TESS) and Kepler missions, along with ground-based transit surveys, have provided a bounty of high-quality targets for Pandora. The final target list for Pandora will not be decided until approximately 6 months before launch in order to capitalize on recent discoveries, but we have developed a notional target list to use in developing a design reference mission. There are more than 20 suitable targets for our notional target list that meet our SNR requirements, of which 16 are already planned for JWST observations. Targets are chosen to maximize both the range of stars and planets probed and the observing efficiency. Our final observing plan will be posted publicly, in advance, to support ground-based observers.

\subsection{Auxiliary Science}
Pandora's legacy will be a unique catalog of long-duration time-series photometry and simultaneous spectroscopy (both during and out of transit) that will enable science beyond Pandora's objectives. 
Furthermore, the baseline science operation includes schedule margin which, if unused, will be available to collect additional data to enhance the science output from the mission and support other ground and space-based observatories. Auxiliary science cases the Pandora team has identified include measuring transit timing variations, measuring the Rayleigh scattering slope of exoplanetary atmospheres, and contemporaneous stellar jitter measurements during ground-based radial velocity observations.


\section{Pandora Mission Overview}
Pandora provides big science in a small payload. The mission is designed to enable long-observing baselines on single targets, precise attitude control, and thermal stability. During the one-year science operations, Pandora will observe over 200 transits---10 transits for each of the 20 targets with $>$12 hours of observing baseline per transit. The two tall poles for the mission are maintaining precise pointing stability and tight thermal control of the NIR sensor. Pandora requires pointing of $<3$ arcseconds over 60 seconds. Pandora will meet this requirement by utilizing high-quality star trackers and including the payload visible camera in the pointing control loop. The NIR sensor will be kept colder than 130 K, with 10 mK temperature stability maintained. This stability will be achieved through the use of a cryocooler and additional thermal control electronics. Pandora is designed to fly as a ride-share and accommodated on an ESPA Grande ring.

\subsection{Concept of Operations}
Pandora is designed for long stares at exoplanet host stars. The observatory needs to be in an orbit that allows for long-baseline observations of single targets, facilitates relatively high data rates, provides a reasonably stable thermal and power environment, and ideally has many opportunities for ride-shares. This leads to Sun-synchronous low-Earth orbit with a crossing time of approximately 6am/6pm (Figure~\ref{fig:orbit}). In this orbit, the whole sky is accessible during the 12 months of science operations, and most regions of the sky are accessible for over 100 days per year. Pandora's top 20 notional targets are highly accessible (enough to observe at least 10 transits per planet and at least 120 hours total per star). During science operations, observation sequences will be uploaded to the spacecraft up to twice weekly. Pandora will be assigned a single target for approximately 24 hrs before moving to the next target (the transit need not fall into the center of the 24-hr baseline). During this 24-hour period, Pandora will collect science data for at least 12 hours with breaks during data down-link, Earth-occultation, and Moon avoidance. For short-orbital-period planets, we may opt to stare at the targets for multiple days. Several target stars host multiple transiting planets; these stars will benefit from increased observing baselines. 

\begin{Figure}
\centering
\includegraphics[width=\textwidth]{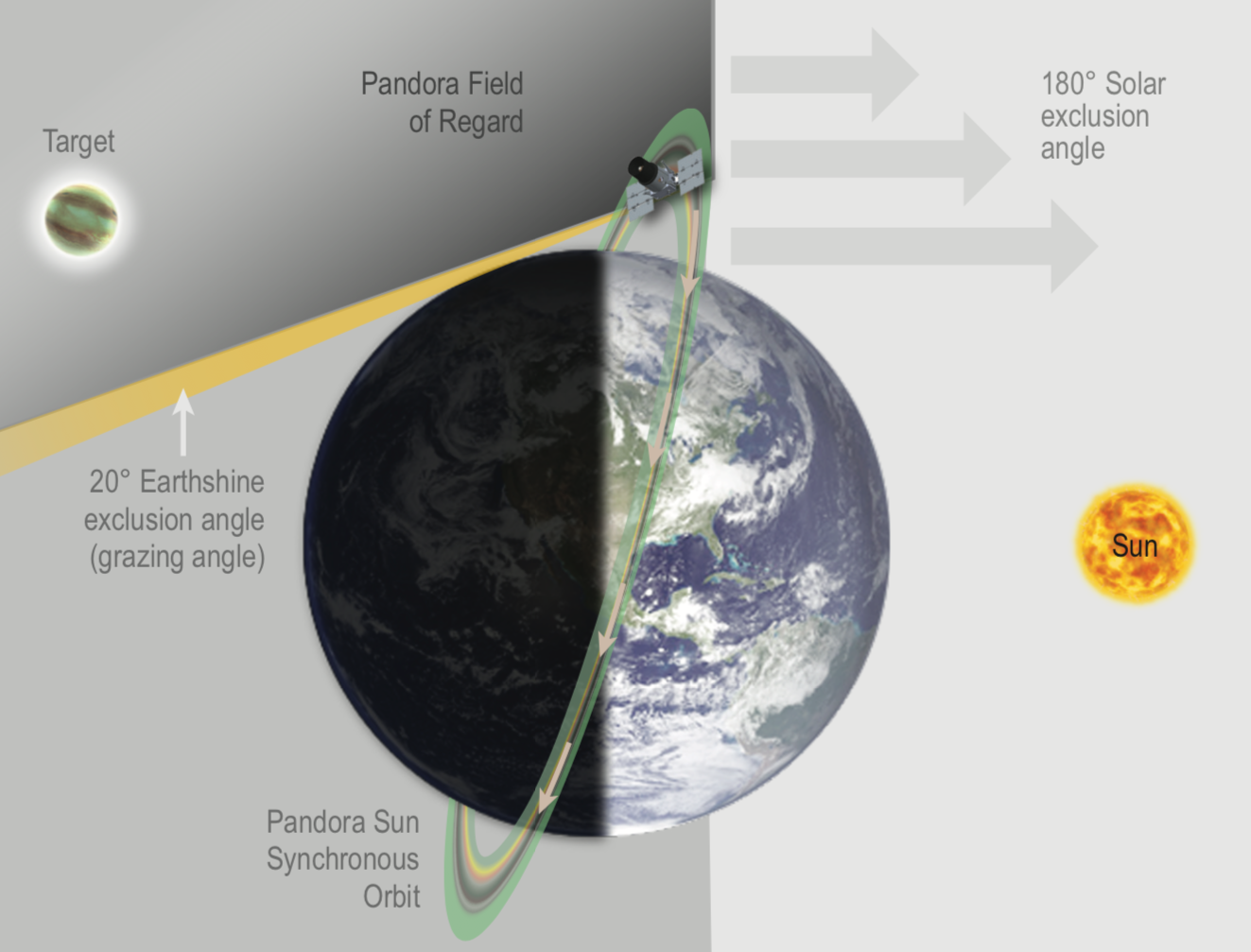}
\captionof{figure}{Pandora's Sun-Synchronous Orbit (SSO) in Low Earth Orbit (LEO) allows long-baseline observations of single targets. The SSO provides a relatively stable thermal environment and excellent ground station access.}
\label{fig:orbit}
\end{Figure}


The Pandora vehicle will be power-positive for most observations, but will require battery power during seasonal eclipses and when science targets are located closer to the Sun (e.g., the telescope points into and out of the page in Figure~\ref{fig:orbit}). Most observations will be performed with the space vehicle in a fixed inertial pointing, although there are some targets during eclipse season that will require the vehicle to perform 180-degree rotations every half-orbit in order to maintain thermal stability.

Pandora's observing timeline is designed to be highly flexible and can be customized to support observations from other space- and ground-based telescopes. Data will be down-linked over X-band approximately every 12 hrs, and calibrated by the science data pipeline, with high-level data products created for public release at regular intervals.

\subsection{Payload}



\begin{Figure}
\centering
\includegraphics[width=0.8\textwidth]{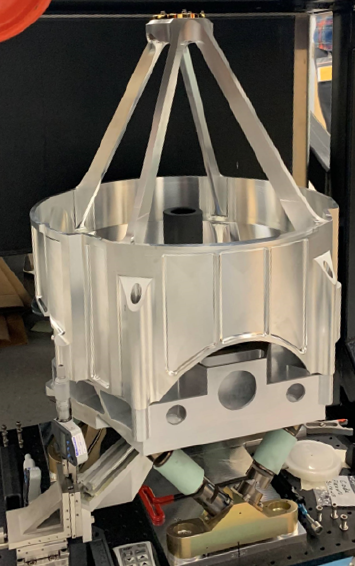}
\captionof{figure}{Pandora uses an all-aluminum 0.45-m Cassegrain telescope developed by Lawrence Livermore National Laboratory (LLNL). Pandora's telescope provides the necessary aperture and stability for characterizing exoplanet atmospheres. }
\label{fig:payload2}
\end{Figure}

The Pandora payload is built around an all-aluminum 0.45-m Cassegrain telescope design (Figure~\ref{fig:payload2}). The telescope is designed to address the challenge of procuring high-performing imaging space systems in a ride-sharing form-factor that are low cost and have rapid acquisition. To help address these challenges, the telescope design minimizes space-vehicle dependencies and maximizes customizability of the aft optics to accommodate a variety of optical configurations. Compared to traditional glass optics, aluminum telescopes are more suited to large-scale manufacturing and batch production. Combined with mechanical interfaces that are machined in parallel with optical surfaces, the telescope has a significant cost- and schedule- advantage over other designs.

The optical assembly is composed of six primary sub-systems: primary mirror assembly, secondary mirror assembly, support spider, relay assembly, optical bench, and mounting bipods. The telescope interfaces with the spacecraft bus through 3 sets of bipod mounts. A qualification unit of the telescope has been fabricated and demonstrated the capability to converge on key performance metrics (Figure~\ref{fig:payload2}). This includes predicted integrated system-level performance of 30 micron RMS spot diameter at 1.1" angular resolution, with optical component surface roughness less than 20 Å.


\begin{Figure}
\centering
\includegraphics[width=0.9\textwidth]{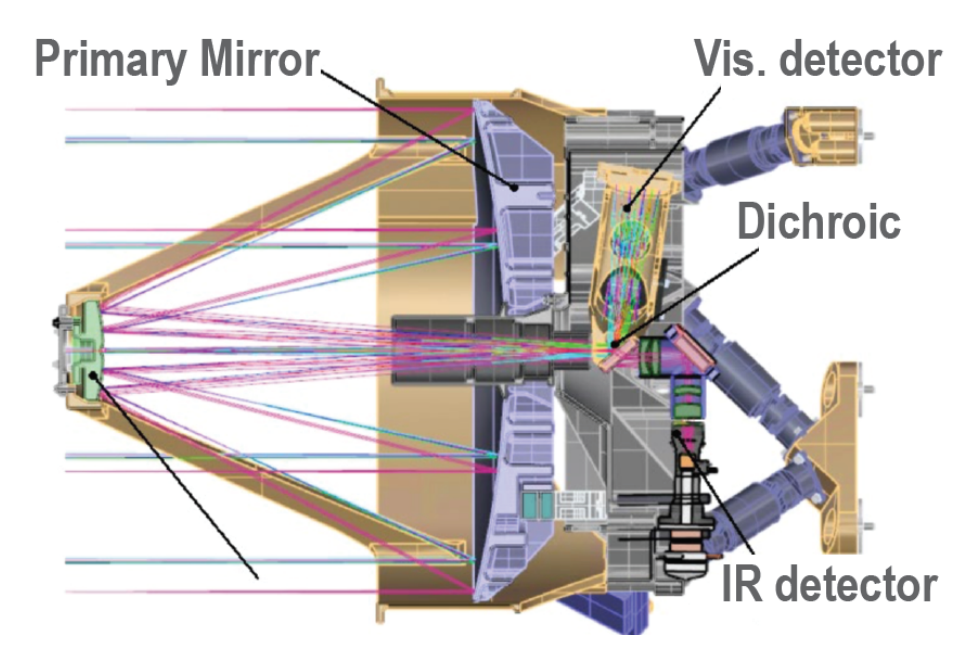}
\captionof{figure}{Pandora's optical system fulfills the sensitivity requirements and provides simultaneous multiwavelength capabilities in a compact, low-complexity design. Visible light is reflected off a dichroic coating on the front surface of the prism, and a focusing parabola brings the visible light to the visible channel detector. NIR light is dispersed through a prism to the infrared detector. Internal and external baffles (not shown) provide off-axis light supression. }
\label{fig:payload3}
\end{Figure}

Pandora's telescope accommodates simultaneous visible channel photometry (400-650 nm) and near-infrared spectroscopy (1000-1600 nm). The aft optical relay system of Pandora is arranged in a plane perpendicular to the optical axis of the telescope and resides behind the primary mirror (Figure~\ref{fig:payload3}). Light is relayed to a dichroic mirror that directs visible light toward the visible imaging sensor, and the NIR light through a full field dispersive element and to the NIR channel sensor chip assembly.

The visible channel has the dual role of providing photometric monitoring of the target star and providing precise pointing information to the attitude control system. The NIR channel has a spectral resolution ($\Delta\lambda/\lambda$) of $>$30 from 1.0-1.6 $\mu$m and uses a HAWAII-2RG sensor from Teledyne Imaging. A precision thermal control system cools the NIR channel to below 130K and holds sub 10mK stability on both the visible and NIR sensors. 

\subsection{Mission Timeline}
\begin{figure*}
\centering
\includegraphics[width=\textwidth]{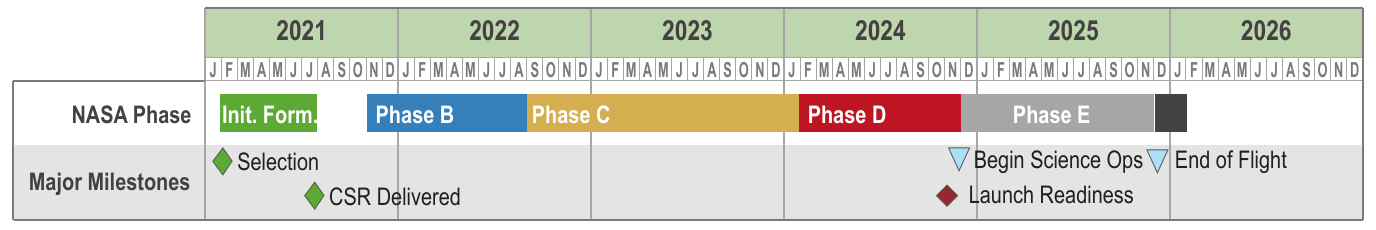}
\captionof{figure}{The Pandora mission will complete the science objectives proposed, archive the data, and publish results within a 5-year timeline. Launch would occur in the mid-2020s, although the precise timeline is dependent on a ride-sharing agreement.}
\label{fig:missiontimeline}
\end{figure*}

Pandora has a 5-year performance period from selection to closeout as shown in Figure~\ref{fig:missiontimeline}. For planning purposes, we have scheduled Pandora in ``Phases'' based on NASA's mission lifecycle terminology. Following the six-month formulation phase and authority to proceed to implementation, Pandora will  start preliminary design and technology completion (Phase B). Final design and fabrication is next (Phase C). System assembly, integration and testing, and launch (Phase D) will follow. The launch readiness is projected to be in late 2024. Operations and science collection (Phase E) will last 1 year, followed by a close-out period.

\subsection{Data Processing and Archiving}
A core philosophy of Pandora is to ensure that the data collected and the tools developed can be valuable beyond the core science team and science objectives. In addition to generating and distributing key science results, our plan will provide the Pandora data and software used by the mission team to the science community in easily accessible formats and venues. Pandora's data products include raw data, calibrated images, calibrated light curves, calibrated spectra, the data processing pipeline, and key mission publications.  All of these products will be archived and publicly distributed in a timely manner.

Pandora's Data Processing Center (DPC) will be located at NASA ARC. They will be responsible for developing and operating the Pandora data analysis pipeline. The pipeline design and implementation will be based on ARC's extensive experience with NASA's exoplanet missions Kepler, K2, and TESS.  

The Infrared Processing and Analysis Center (IPAC) at Caltech will archive, curate, and serve the final Level 1--3 data products, including the pixel data, light curves, and transmission spectra from the Pandora mission. The data will be served through standard interfaces at NASA Exoplanet Science Institute (NExScI), placing them alongside and in context with other NASA data sets, and enhancing their value to archival researchers and future astronomical mission planners.












\section{Pandora Management}
Pandora is a collaboration among GSFC, which is the PI institution, LLNL, which will manage the mission, and Co-Is from ARC, IPAC, the University of Arizona, and MIT. Pandora's management organization includes team members with decades of space mission experience, and early career scientists and engineers who serve in leadership roles for the detectors, operations, data processing, and science output. The Pandora team uses a mentoring model that matches experienced team members with early career team members, thereby increasing the pool of diverse future spaceflight leaders. Pandora has also developed shadow positions, which enables a graduate student to shadow the Project Scientist for the first program year (and opportunities for new graduate student shadows in subsequent program years).
 
Pandora has a Science Working Group (SWG) structure led by the Project Scientist and involving Pandora Co-Investigators and Collaborators. The SWGs include an Exoplanets SWG, a Stellar Contamination SWG, a Data Analysis SWG, a Target Selection and Observing Strategy SWG, and an Auxiliary Science SWG, with significant overlap of team members. The SWGs will also help facilitate external ground- and space-based complementary observations of Pandora's targets that---while not necessary for fulfilling Pandora's objectives---will maximize Pandora's science output. Opportunities for community engagement and participation are in development.

\section{Conclusions}
Pandora science cannot be done with current or upcoming ground- or space-based facilities that will operate in the JWST era. Detecting water absorption in transmission spectra is more challenging from the ground than from space, because Earth’s atmosphere has poor transmission and is highly variable at these wavelengths. While HST has wavelength bands in the visible and IR, HST does not perform \textit{simultaneous} optical and IR observations. JWST does not observe in the optical ($<0.6$\,micron), which is critical for quantifying stellar contamination. A combination of ground-based photometric surveys operating simultaneously with HST or JWST cannot do Pandora science either. The precision of ground-based data limit inferences of spot properties \cite{Rosich2020}, and obtaining long time-baseline observations is not possible with HST/JWST due to the highly oversubscribed nature of these telescopes. To summarize, Pandora’s long-time-baseline, simultaneous optical and NIR data are unique and fill a critical gap in NASA's goal of characterizing the atmospheres of small exoplanets.

\begin{Figure}
\centering
\includegraphics[width=\textwidth]{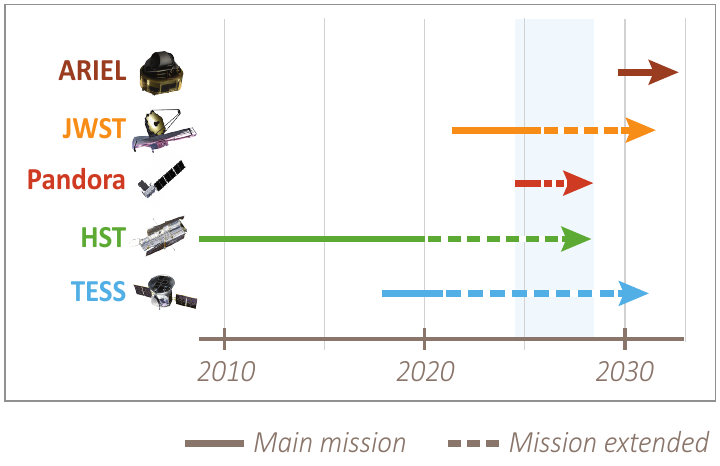}
\captionof{figure}{Pandora will overlap in time with JWST's prime mission and potentially with HST, and will inform target selection for JWST.}
\label{fig:joint-timelines}
\end{Figure}

Pandora is expected to operate concurrently with JWST (Figure \ref{fig:joint-timelines}). JWST has four science instruments with modes for observing transiting exoplanets to high precision and longer IR wavelengths; however, stellar contamination will persist for JWST transmission spectra \cite{Iyer2020}. We anticipate the opportunity for simultaneous observations in which JWST will do short-baseline deep dives and Pandora will provide long-baseline and visible-band observations. Pandora will further maximize JWST science by informing the interpretation of data collected prior to Pandora, revisiting high-profile targets observed by JWST, and informing target selection for observations following JWST GO Cycle 5 (assuming an Oct 2021 JWST launch). Pandora will determine whether concurrent visible photometry is always required for accurate transmission spectroscopy of planetary atmospheres and will indicate if certain types of stars and planets are not likely to suffer from this type of stellar contamination. Looking ahead, the European Space Agency's Atmospheric Remote-sensing Infrared Exoplanet Large-survey (Ariel) mission has a NASA component, the Contribution to ARIEL Spectroscopy of Exoplanets (CASE) \cite{Zellem2019}. By selecting Pandora now, we ensure that Pandora is operating simultaneously with JWST and can make contributions to observing strategies for future exoplanet missions like Ariel/CASE.

A successful Pandora mission will demonstrate the value of SmallSats for performing high-impact science that can complement flagship missions at a low cost and rapid timescale. Moreover, SmallSats are an excellent platform for training the next generation of leaders in space mission development and management.

\subsection*{Acknowledgments}


Pandora is supported by NASA's Astrophysics Pioneers Program. Pre-proposal work was supported by GSFC's Internal Research and Development Program and a Mission Planning Lab at Wallops Flight Facility. We are indebted to the many, many people who contributed to the various iterations of Pandora.






\bibliography{pandora.bib}

\begin{thebibliography}{10}

\bibitem{Seager2000}
S.~{Seager} and D.~D. {Sasselov}.
\newblock {Theoretical Transmission Spectra during Extrasolar Giant Planet
  Transits}.
\newblock {\em \apj}, 537(2):916--921, July 2000.

\bibitem{Sing2016}
David~K. {Sing}, Jonathan~J. {Fortney}, Nikolay {Nikolov}, Hannah~R.
  {Wakeford}, Tiffany {Kataria}, Thomas~M. {Evans}, Suzanne {Aigrain}, Gilda~E.
  {Ballester}, Adam~S. {Burrows}, Drake {Deming}, Jean-Michel {D{\'e}sert},
  Neale~P. {Gibson}, Gregory~W. {Henry}, Catherine~M. {Huitson}, Heather~A.
  {Knutson}, Alain {Lecavelier Des Etangs}, Frederic {Pont}, Adam~P. {Showman},
  Alfred {Vidal-Madjar}, Michael~H. {Williamson}, and Paul~A. {Wilson}.
\newblock {A continuum from clear to cloudy hot-Jupiter exoplanets without
  primordial water depletion}.
\newblock {\em \nat}, 529(7584):59--62, January 2016.

\bibitem{Greene2016}
Thomas~P. {Greene}, Michael~R. {Line}, Cezar {Montero}, Jonathan~J. {Fortney},
  Jacob {Lustig-Yaeger}, and Kyle {Luther}.
\newblock {Characterizing Transiting Exoplanet Atmospheres with JWST}.
\newblock {\em \apj}, 817(1):17, January 2016.

\bibitem{Greene2019}
Thomas {Greene}, Natalie {Batalha}, Jacob {Bean}, Thomas {Beatty}, Jeroen
  {Bouwman}, Jonathan {Fortney}, Yasuhiro {Hasegawa}, Thomas {Henning}, David
  {Lafreniere}, Pierre-Olivier {Lagage}, George {Rieke}, Thomas {Roellig},
  Everett {Schlawin}, and Kevin {Stevenson}.
\newblock {Characterizing Transiting Exoplanets with JWST Guaranteed Time and
  ERS Observations}.
\newblock {\em \baas}, 51(3):61, May 2019.

\bibitem{Baliunas1995}
S.~L. {Baliunas}, R.~A. {Donahue}, W.~H. {Soon}, J.~H. {Horne}, J.~{Frazer},
  L.~{Woodard-Eklund}, M.~{Bradford}, L.~M. {Rao}, O.~C. {Wilson}, Q.~{Zhang},
  W.~{Bennett}, J.~{Briggs}, S.~M. {Carroll}, D.~K. {Duncan}, D.~{Figueroa},
  H.~H. {Lanning}, T.~{Misch}, J.~{Mueller}, R.~W. {Noyes}, D.~{Poppe}, A.~C.
  {Porter}, C.~R. {Robinson}, J.~{Russell}, J.~C. {Shelton}, T.~{Soyumer},
  A.~H. {Vaughan}, and J.~H. {Whitney}.
\newblock {Chromospheric Variations in Main-Sequence Stars. II.}
\newblock {\em \apj}, 438:269, January 1995.

\bibitem{Dittmann2009}
Jason~A. {Dittmann}, Laird~M. {Close}, Elizabeth~M. {Green}, and Mike
  {Fenwick}.
\newblock {A Tentative Detection of a Starspot During Consecutive Transits of
  an Extrasolar Planet from the Ground: No Evidence of a Double Transiting
  Planet System Around TrES-1}.
\newblock {\em \apj}, 701(1):756--763, August 2009.

\bibitem{Sing2011}
D.~K. {Sing}, F.~{Pont}, S.~{Aigrain}, D.~{Charbonneau}, J.~M. {D{\'e}sert},
  N.~{Gibson}, R.~{Gilliland}, W.~{Hayek}, G.~{Henry}, H.~{Knutson},
  A.~{Lecavelier Des Etangs}, T.~{Mazeh}, and A.~{Shporer}.
\newblock {Hubble Space Telescope transmission spectroscopy of the exoplanet HD
  189733b: high-altitude atmospheric haze in the optical and near-ultraviolet
  with STIS}.
\newblock {\em MNRAS}, 416(2):1443--1455, September 2011.

\bibitem{Deming2011}
Drake {Deming}, Pedro~V. {Sada}, Brian {Jackson}, Steven~W. {Peterson}, Eric
  {Agol}, Heather~A. {Knutson}, Donald~E. {Jennings}, Flynn {Haase}, and Kevin
  {Bays}.
\newblock {Kepler and Ground-based Transits of the Exo-Neptune HAT-P-11b}.
\newblock {\em \apj}, 740(1):33, October 2011.

\bibitem{Sanchis2011}
Roberto {Sanchis-Ojeda} and Joshua~N. {Winn}.
\newblock {Starspots, Spin-Orbit Misalignment, and Active Latitudes in the
  HAT-P-11 Exoplanetary System}.
\newblock {\em \apj}, 743(1):61, December 2011.

\bibitem{Morris2017}
Brett~M. {Morris}, Leslie {Hebb}, James R.~A. {Davenport}, Graeme {Rohn}, and
  Suzanne~L. {Hawley}.
\newblock {The Starspots of HAT-P-11: Evidence for a Solar-like Dynamo}.
\newblock {\em \apj}, 846(2):99, September 2017.

\bibitem{Rackham2018}
B.~V. {Rackham}, D.~{Apai}, and M.~S. {Giampapa}.
\newblock {The Transit Light Source Effect: False Spectral Features and
  Incorrect Densities for M-dwarf Transiting Planets}.
\newblock {\em \apj}, 853:122, February 2018.

\bibitem{McCullough2014}
P.~R. {McCullough}, N.~{Crouzet}, D.~{Deming}, and N.~{Madhusudhan}.
\newblock {Water Vapor in the Spectrum of the Extrasolar Planet HD 189733b. I.
  The Transit}.
\newblock {\em \apj}, 791(1):55, August 2014.

\bibitem{Rackham2017}
Benjamin {Rackham}, N{\'e}stor {Espinoza}, D{\'a}niel {Apai}, Mercedes
  {L{\'o}pez-Morales}, Andr{\'e}s {Jord{\'a}n}, David~J. {Osip}, Nikole~K.
  {Lewis}, Florian {Rodler}, Jonathan~D. {Fraine}, Caroline~V. {Morley}, and
  Jonathan~J. {Fortney}.
\newblock {ACCESS I: An Optical Transmission Spectrum of GJ 1214b Reveals a
  Heterogeneous Stellar Photosphere}.
\newblock {\em \apj}, 834(2):151, January 2017.

\bibitem{Apai2018}
D.~{Apai}, B.~V. {Rackham}, M.~S. {Giampapa}, D.~{Angerhausen}, J.~{Teske},
  J.~{Barstow}, L.~{Carone}, H.~{Cegla}, S.~D. {Domagal-Goldman},
  N.~{Espinoza}, H.~{Giles}, M.~{Gully-Santiago}, R.~{Haywood}, R.~{Hu},
  A.~{Jordan}, L.~{Kreidberg}, M.~{Line}, J.~{Llama}, M.~{L{\'o}pez-Morales},
  M.~S. {Marley}, and J.~{de Wit}.
\newblock {Understanding Stellar Contamination in Exoplanet Transmission
  Spectra as an Essential Step in Small Planet Characterization}.
\newblock {\em ArXiv e-prints}, March 2018.

\bibitem{Espinoza2019}
N{\'e}stor {Espinoza}, Benjamin~V. {Rackham}, Andr{\'e}s {Jord{\'a}n},
  D{\'a}niel {Apai}, Mercedes {L{\'o}pez-Morales}, David~J. {Osip}, Simon~L.
  {Grimm}, Jens {Hoeijmakers}, Paul~A. {Wilson}, Alex {Bixel}, Chima
  {McGruder}, Florian {Rodler}, Ian {Weaver}, Nikole~K. {Lewis}, Jonathan~J.
  {Fortney}, and Jonathan {Fraine}.
\newblock {ACCESS: a featureless optical transmission spectrum for WASP-19b
  from Magellan/IMACS}.
\newblock {\em MNRAS}, 482(2):2065--2087, January 2019.

\bibitem{Rackham2019b}
Benjamin {Rackham}, Arazi {Pinhas}, D{\'a}niel {Apai}, Rapha{\"e}lle {Haywood},
  Heather {Cegla}, N{\'e}stor {Espinoza}, Johanna {Teske}, Michael
  {Gully-Santiago}, Gioia {Rau}, Brett~M. {Morris}, Daniel {Angerhausen},
  Thomas {Barclay}, Ludmila {Carone}, P.~Wilson {Cauley}, Julien {de Wit},
  Shawn {Domagal-Goldman}, Chuanfei {Dong}, Diana {Dragomir}, Mark~S.
  {Giampapa}, Yasuhiro {Hasegawa}, Natalie~R. {Hinkel}, Renyu {Hu}, Andr{\'e}s
  {Jord{\'a}n}, Irina {Kitiashvili}, Laura {Kreidberg}, Carey {Lisse}, Joe
  {Llama}, Mercedes {L{\'o}pez-Morales}, Bertrand {Mennesson}, Karan
  {Molaverdikhani}, David~J. {Osip}, and Elisa~V. {Quintana}.
\newblock {Constraining Stellar Photospheres as an Essential Step for
  Transmission Spectroscopy of Small Exoplanets}.
\newblock {\em Bulletin of the AAS}, 51(3):328, May 2019.

\bibitem{Rackham2019a}
Benjamin~V. {Rackham}, D{\'a}niel {Apai}, and Mark~S. {Giampapa}.
\newblock {The Transit Light Source Effect. II. The Impact of Stellar
  Heterogeneity on Transmission Spectra of Planets Orbiting Broadly Sun-like
  Stars}.
\newblock {\em \aj}, 157(3):96, March 2019.

\bibitem{Wakeford2019}
H.~R. {Wakeford}, N.~K. {Lewis}, J.~{Fowler}, G.~{Bruno}, T.~J. {Wilson}, S.~E.
  {Moran}, J.~{Valenti}, N.~E. {Batalha}, J.~{Filippazzo}, V.~{Bourrier}, S.~M.
  {H{\"o}rst}, S.~M. {Lederer}, and J.~{de Wit}.
\newblock {Disentangling the Planet from the Star in Late-Type M Dwarfs: A Case
  Study of TRAPPIST-1g}.
\newblock {\em \aj}, 157(1):11, January 2019.

\bibitem{Pinhas2018}
Arazi {Pinhas}, Benjamin~V. {Rackham}, Nikku {Madhusudhan}, and D{\'a}niel
  {Apai}.
\newblock {Retrieval of planetary and stellar properties in transmission
  spectroscopy with AURA}.
\newblock {\em MNRAS}, 480(4):5314--5331, November 2018.

\bibitem{Zhang2018}
Z.~{Zhang}, Y.~{Zhou}, B.~{Rackham}, and D.~{Apai}.
\newblock {The Near-Infrared Transmission Spectra of TRAPPIST-1 Planets b, c,
  d, e, f, and g and Stellar Contamination in Multi-Epoch Transit Spectra}.
\newblock {\em ArXiv e-prints}, February 2018.

\bibitem{Pont2008}
F.~{Pont}, H.~{Knutson}, R.~L. {Gilliland}, C.~{Moutou}, and D.~{Charbonneau}.
\newblock {Detection of atmospheric haze on an extrasolar planet: the 0.55-1.05
  {$\mu$}m transmission spectrum of HD 189733b with the HubbleSpaceTelescope}.
\newblock {\em MNRAS}, 385:109--118, March 2008.

\bibitem{Pont2013}
F.~{Pont}, D.~K. {Sing}, N.~P. {Gibson}, S.~{Aigrain}, G.~{Henry}, and
  N.~{Husnoo}.
\newblock {The prevalence of dust on the exoplanet HD 189733b from Hubble and
  Spitzer observations}.
\newblock {\em MNRAS}, 432(4):2917--2944, July 2013.

\bibitem{Berta2012}
Zachory~K. {Berta}, David {Charbonneau}, Jean-Michel {D{\'e}sert}, Eliza
  {Miller-Ricci Kempton}, Peter~R. {McCullough}, Christopher~J. {Burke},
  Jonathan~J. {Fortney}, Jonathan {Irwin}, Philip {Nutzman}, and Derek
  {Homeier}.
\newblock {The Flat Transmission Spectrum of the Super-Earth GJ1214b from Wide
  Field Camera 3 on the Hubble Space Telescope}.
\newblock {\em \apj}, 747(1):35, March 2012.

\bibitem{Deming2017}
L.~Drake Deming and Sara Seager.
\newblock Illusion and reality in the atmospheres of exoplanets.
\newblock {\em Journal of Geophysical Research: Planets}, 122(1):53--75, 2017.

\bibitem{NAP25187}
{National Academies of Sciences, Engineering, and Medicine}.
\newblock {\em Exoplanet Science Strategy}.
\newblock The National Academies Press, Washington, DC, 2018.

\bibitem{Kowalski2019}
Adam {Kowalski}, Karel {Schrijver}, Valentin {Pillet}, and Serena {Criscuoli}.
\newblock {Developing a vision for exoplanetary transit spectroscopy: a shared
  window on the analysis of planetary atmospheres and of stellar magnetic
  structure}.
\newblock {\em \baas}, 51(3):149, May 2019.

\bibitem{Rosich2020}
A.~{Rosich}, E.~{Herrero}, M.~{Mallonn}, I.~{Ribas}, J.~C. {Morales},
  M.~{Perger}, G.~{Anglada-Escud{\'e}}, and T.~{Granzer}.
\newblock {Correcting for chromatic stellar activity effects in transits with
  multiband photometric monitoring: application to WASP-52}.
\newblock {\em \aap}, 641:A82, September 2020.

\bibitem{Iyer2020}
Aishwarya~R. {Iyer} and Michael~R. {Line}.
\newblock {The Influence of Stellar Contamination on the Interpretation of
  Near-infrared Transmission Spectra of Sub-Neptune Worlds around M-dwarfs}.
\newblock {\em \apj}, 889(2):78, February 2020.

\bibitem{Zellem2019}
Robert~T. {Zellem}, Mark~R. {Swain}, Nicolas~B. {Cowan}, Geoffrey {Bryden},
  Thaddeus~D. {Komacek}, Mark {Colavita}, David {Ardila}, Gael~M. {Roudier},
  Jonathan~J. {Fortney}, Jacob {Bean}, Michael~R. {Line}, Caitlin~A.
  {Griffith}, Evgenya~L. {Shkolnik}, Laura {Kreidberg}, Julianne~I. {Moses},
  Adam~P. {Showman}, Kevin~B. {Stevenson}, Andre {Wong}, John~W. {Chapman},
  David~R. {Ciardi}, Andrew~W. {Howard}, Tiffany {Kataria}, Eliza M.~R.
  {Kempton}, David {Latham}, Suvrath {Mahadevan}, Jorge {Mel{\'e}ndez}, and
  Vivien {Parmentier}.
\newblock {Constraining Exoplanet Metallicities and Aerosols with the
  Contribution to ARIEL Spectroscopy of Exoplanets (CASE)}.
\newblock {\em \pasp}, 131(1003):094401, September 2019.

\end{thebibliography}
\bibliographystyle{unsrt}

\end{multicols*}
\end{document}